# Impact of a reclassification of Web of Science articles on bibliometric indicators

**Agénor Lahatte[1] (1), Élisabeth de Turckheim[2] (1)**

(1) Hcéres - Science and Technology Observatory

19, rue Poissonnière  75002 PARIS

Corresponding author: agenor.lahatte@hceres.fr

**Abstract**

This work aims at evaluating a reclassification of Web of Science articles implemented at OST. Articles from the 254 scientific categories of the Web of Science were reclassified at article level in 242 modified categories and 11 disciplines using the method of S. Milojević (2020). The reclassification is based on paper references categories and it no longer assigns papers to multiple or to multidisciplinary categories. It improves the accuracy and the modularity of the WoS classification. As there are important changes in document assignment at the lowest level, usual indicators such as disciplinary profiles or field normalized indicators are significantly modified. This study examines some of these modifications to provide explanations for the recipients of OST reports. Changes in specialization indexes reveal specific journal choices by scientists. In a sample of 25 countries, Brazil and China offer examples of facilities or constraints for selecting journals to publish certain research works.

## Introduction

Scientific classifications are necessary for bibliometric analyses and many classification schemes have been proposed and used. Historically, scientific publications were classified through journals. The classifications of WoS and Scopus databases have been challenged (Wang & Waltman, 2016) and new journal classifications have been proposed, using quantitative methods such as similarity measures of citation data or hybrid methods (Borner et al. 2012, Leydesdorff et al. 2017, Archambault et al. 2011). Multidisciplinary journals and journals with a lower specialization that does not fit with a precise classification scheme are hard to classify, however. These journals are usually assigned to several categories or to multidisciplinary categories. The use of multiple assignments leads to computational complications due to fractional disciplinary counts in indicator definitions. In addition, designing and implementing precise statistical methods that rely on sample simulations, like those developed by Thelwall and Fairclough (2017), is becoming more challenging.

Classifications at the level of individual publications avoid these disadvantages (Boyack et al. 2011; Klavans and Boyack 2017; Waltman and van Eck, 2012). High quality algorithms have been developed (Traag et al, 2019). The CWTS classification for instance is based on direct citation data and open algorithms (Waltman & van Eck, 2012) and is now based on data of the OpenAlex database (Waltman & van Eck, 2024).

The French Science and Technology Observatory (OST) uses the WoS scientific categories as the basic level of its scientific nomenclature. To shift to a paper-based classification and to keep the frame of this scientific taxonomy, we implemented the reclassification algorithm of papers proposed by Milojević (2020), which assigns a single WoS category to each publication and reclassifies publications of multidisciplinary journals into disciplinary categories. The consistency between the category of a paper and of its references is enhanced by this reclassification. But changes of category assignment are numerous and this leads to significant changes in the usual bibliometric indicators. Shu et al. (2020) exhibit how their paper-level reclassification affects the productivity rankings of actors within disciplines.

As OST produces regular reports for French institutions, these modifications require a thorough understanding of the changes in standard bibliometric indicators that result from this reclassification. The paper aims to examine the changes in two indicators, the specialization index and the Mean Normalized Citation Score (MNCS). We examine the case of the 25 most productive countries on the period 2010-2022.

Section 1 describes the OST adaptation of Milojević's reclassification algorithm and how it is extended to a higher hierarchical level. In Section 2, modified categories are surveyed to identify dispersed and attractive categories with similar statistics to Shu et al. (2020). Section 3 shows how the distribution of world publications into disciplines is modified and how countries specialization indexes are related to migration rates between disciplines. Section 4 briefly examines the changes in countries MNCS, using the example of MNCS in Mathematics.

---

1   https://orcid.org/0000-0003-1031-4364
2   https://orcid.org/0000-0003-2372-5410

# 1 Reclassification of WoS documents

Stasa Milojević (20120) suggested to keep nominal WoS categories and to revise the assignment of each article in these categories. Milojević's reclassification is based on citation information, selecting their most frequent WoS category of the references of a paper as its paper-category *(P-category)*. A unique category is chosen for each article. Papers in multidisciplinary journals are assigned to disciplinary categories.

## 1.1 Data

We use the whole in-house version of the WOS database[3] as uploaded on May $5^{th}$, 2023. For the classification algorithm, we use all documents having WoS categories (called J-categories, as they are defined by the journal). Whereas all possible document in the database is used in the classification algorithm, we provide statistics only for the "standard perimeter" of documents that are of type Article, Review or Conference Proceedings (denoted *Corpus Y*).

## 1.2 Reclassification algorithm

Each paper in the database with at least 2 references in the base will be assigned a unique P-category (category assigned by paper) obtained by a reclassification algorithm very similar to Milojević's (20120). The P-categories have the same titles as the original WoS J-categories. Ten multidisciplinary categories are removed, as there are not selected as possible P-categories[4].

The algorithm selects a unique category for each paper which is the most frequent category of its references. WoS categories of references are counted with fractional counts and references in the 10 multidisciplinary categories are ignored. A first run allows to assign a P-category to the documents that have a unique maximal J-category. Two others runs are necessary to adjust the P-category of a paper to the P-categories of its references when they are modified. This closes the first step (Run 1, 2 and 3, Table 1). As Milojević (2020), we solve the last tied issues in a second step, adding the paper category to the category count (Run 4, Table 1) and, when ties remain, selecting the largest WoS category (Run 5, Table 1).

**Table 1.** Counts of reclassified documents at the different steps of the reclassification algorithm in corpus Z the subset of documents with at least 2 references. Counts are for Article, Review and Conference Proceedings document types

|  | Documents to classify | Documents with assigned P-category | Documents without assigned P-category | Non-assigned documents with tied references | Non assigned documents with multiD references |
|---|---|---|---|---|---|
| *Step1* | | | | | |
| Run 1 | 36,908,770 | 33,848,535 | 3,060,235 | 3,012,765 | 47,470 |
| Run 2 | 36,908,770 | 34,482,274 | 2,426,496 | 2,406,552 | 19,944 |
| Run 3 | 36,908,770 | 34,676,784 | *2,231,986* | 2,214,203 | 17,783 |
| *Step 2* | | | | | |
| Run 4 | *2,231,986* | 997,342 | *1,234,644* | 1,224,853 | 9,791 |
| Run 5 | *1,234,644* | 1,224,853 | 9,791 | 0 | 9,791 |
| *Steps 1 + 2* | 36,908,770 | 36,898,979 | 9,791 | | |

A rule is finally defined to choose the P-category of 9,791 papers in multidisciplinary J-categories that have all their references in multidisciplinary J-categories. We assign these documents to the largest P-category of all reclassified papers from the same J-multidisciplinary category.

## 1.3 Aggregating categories into disciplines

OST uses a higher level of classification, that consists in 11 disciplines[5] defined as sets of WoS categories - with some categories assigned to two disciplines (Bassecoulard & Zitt, 1999). To revise this level of the classification, we first define a P-discipline for each document with the previous algorithm initialized with the discipline assignment of documents derived from their journal. This step assigns a provisional P-discipline to each document. We assign then choose a single discipline for each P-category which is the largest P-discipline of the documents of the category. In some cases the category is broken down into two or more disciplines. The rule is to break down a category when the three following conditions are satisfied

- the predominant discipline is less than 80% on the category,

---

3  including 5 WoS Indexes: Science Citation Index expanded, Social Sciences Citation, Art & Humanities Citation, Essential Sources Citation Index, Conference Proceedings. The WoS-OST database contains scientific publications since 1999.

- the share of another discipline is more than 15% and has at least 5,000 documents,
- the selected disciplines are from different *domains*[6].

For example, the category SUBSTANCE ABUSE (GM) is broken down into two categories SUBSTANCE ABUSE-Medical Research and SUBSTANCE ABUSE-Social Sciences with codes GM-02 and GM-SS (Table 2).

In these cases, documents of the category are assigned to the sub-category related to their P-discipline if it exists or to the sub-category of a discipline of the same large domain.

**Table 2.** The 11 P-categories broken down into 2 or 3 mono-discipline categories

| CODE | CATEGORY | Disc 1 | Disc 2 | Disc 3 |
|---|---|---|---|---|
| GM | SUBSTANCE ABUSE | 02 | SS | - |
| IG | ENGINEERING, BIOMEDICAL | 07 | 02 | - |
| LJ | GERONTOLOGY | SS | 02 | - |
| NE | PUBLIC, ENVIRONMENTAL & OCCUPATI | 02 | SS | - |
| PI | MARINE & FRESHWATER BIOLOGY | 06 | 03 | - |
| PW | MEDICAL LABORATORY TECHNOLOGY | 02 | 07 | - |
| RZ | NURSING | 02 | SS | - |
| VE | PSYCHIATRY | 02 | SS | - |
| VX | PSYCHOLOGY, EXPERIMENTAL | SS | 01 | - |
| WC | REHABILITATION | 02 | SS | - |
| HB | EDUCATION, SCIENTIFIC DISCIPLINES | 02 | 05 | SS |

We therefore have two classifications of documents into the 11 disciplines: the old OST disciplines derived from WoS categories and from the old correspondence: to be short, we call them *WoS disciplines* and we now call *OST disciplines* the new disciplines based on OST categories and on the new hierarchical correspondence just described (DataSet1, https://zenodo.org/records/15606281).

## *1.4 Final adjustments*

*Documents with less than two references*

In the OST version of the WoS database (May 2023), there are *6,206,034* documents of type Article, Review and Conference Proceeding with less than two references in the base (corpus denoted W). We leave these documents in their J-Category (MULTIDISCIPLINARY categories excluded) - or the largest category if the journal is assigned to multiple categories. The size of the whole corpus Y of documents of type Article, Review and Conference Proceeding (called OST standard perimeter) is

$$\#Y = \#Z + \#W = 36{,}908{,}770 + 6{,}206{,}034 = 43{,}114{,}804$$

*Removing 14 very small categories*

Some P-categories have very few documents. This is the case for some recently introduced categories or for categories of interface fields that are mainly reclassified in one of the historic fields they emerged from. Too small categories may lead to hazardous normalization of individual scores. Therefore these small categories are removed and their documents are assigned to the second most frequent P-category of the whole set of their references (Table 3).

---

[4] AGRICULTURE, MULTIDISCIPLINARY (AH) - HUMANITIES, MULTIDISCIPLINARY (BQ), BIOLOGY (CU), CHEMISTRY, MULTIDISCIPLINARY (DY) - ENGINEERING, MULTIDISCIPLINARY (IF), GEOSCIENCES, MULTIDISCIPLINARY (LE), MATERIALS SCIENCE, MULTIDISCIPLINARY (PM), MULTIDISCIPLINARY SCIENCES (RO), PHYSICS, MULTIDISCIPLINARY (UI), PSYCHOLOGY, MULTIDISCIPLINARY (VJ).
[5] Humanities (SH), Social Science(SS), Biology (01), Medical research (02), Applied Biology & Ecology (03),

Table 3. The 14 small categories merged with another category

| | REMOVED CATEGORY | # in Y (1999-2022) | | MERGED CATEGORY | DISCIPLINES |
|---|---|---|---|---|---|
| RX | NEUROIMAGING | 44 | RT | CLINICAL NEUROLOGY | 02 |
| BV | PSYCHOLOGY, BIOLOGICAL | 302 | CN | BEHAVIORAL SCIENCES | 01 |
| QS | QUANTUM SCIENCE & TECHNOLOGY | 449 | UH | PHYSICS, ATOMIC, MOLECULAR & C | 05 |
| CT | CELL & TISSUE ENGINEERING | 453 | IG02 | ENGINEERING, BIOMEDICAL_02 | 01→02 |
| OO | MEDICAL ETHICS | 505 | PY | MEDICINE, GENERAL & INTERNAL | 02 |
| ML | PRIMARY HEALTH CARE | 1,284 | PY | MEDICINE, GENERAL & INTERNAL | 02 |
| FS | DANCE | 1,358 | YG | THEATER | SH |
| MR | HISTORY OF SOCIAL SCIENCES | 1,408 | MM | HISTORY | SH |
| QL | LOGIC | 1,596 | PQ | MATHEMATICS | 075→08 |
| OU | LIMNOLOGY | 1,629 | JA | ENVIRONMENTAL SCIENCES | 06 |
| AZ | ANDROLOGY | 1,803 | WF | REPRODUCTIVE BIOLOGY | 02→01 |
| WV | SOCIAL SCIENCES, BIOMEDICAL | 2,561 | PY | MEDICINE, GENERAL & INTERNAL | SS→02 |
| PS | SOCIAL SCIENCES, MATHEMATICAL | 2,726 | XY | STATISTICS & PROBABILITY | SS→08 |
| VS | PSYCHOLOGY, MATHEMATICAL | 2,913 | XY | STATISTICS & PROBABILITY | SH→08 |
| | Total | 19,031 | | | |

## 1.5 OST classification: summary

The final 2-level OST new classification is therefore strictly hierarchical with a lower level of 242 P-categories (i.e. 254 WoS categories - 10 MULTIDISCIPLINARY + 12 split categories - 14 removed small categories) and a higher level of 11 disciplines.

After merging the 14 small categories in larger categories, the number of papers that stay in the same nominal category (i.e. with a P-category identical to one of its J-categories) is 54.03% for the corpus Y of all papers in the standard OST perimeter[7] for the period 2010-2022 and 67.41% from the same corpus are reclassified in the same discipline. For these counts, we do not use fractional counts in WoS categories and we consider that a paper reclassified in one of its WoS category does not count as a category change.

This proportion may seem low and suggests that the reclassification process is susceptible to impact usual bibliometric indicators. However, this proportion of changes is similar to that of Chinese authors when they choose the category of their papers, independently from the category of the journal (Shu et al., 2019). For Milojević (2020), 58% of documents with a unique non multidisciplinary WoS category are reclassified in the same category.

Later, we demonstrate that there are significant differences in these proportions between the categories.

## 1.6 Consistency of the new classification

The reclassification of WoS articles is expected to improve the quality of the WoS journal classification. To measure the overall gain of consistency, we use two indicators: modularity and accuracy.

### Modularity

The *modularity* of a classification is the objective function in the Leiden algorithm to detect connected communities in graphs (Traag et al., 2019).

The global modularity H of a classification compares, in each class *C*, the proportion of observed edges to the expected proportion of edges in the random distribution, conditionally to the nodes degrees in the class. Edges between papers are direct citation links.

The formula for this modularity index is $H = \sum_C \frac{1}{2m}(e_C - \frac{K_C^2}{2m})$ where $e_C$ is the number of observed internal edges, $K_C$ is the sum of the degrees of the nodes in the class and $m$ is the number of undirected edges in the whole graph.

We compute the modularity of the new classification on corpus Y (documents of type Article, Review and Conference Proceeding) in the period 2010-2022, excluding documents from the removed multidisciplinary categories.

The comparison between WoS classification and OST reclassification shows a highly significant improvement in modularity. The modularity of the OST classification is much closer to that of meso-level topics of the Leiden Citation

---

Chemistry (04), Physics (05), Earth & Universe sciences (06), Engineering (07), Computer science (75), Mathematics (08)

6 Life sciences, Physics and Engineering, Humanities and Social Science
7 For this count, we do not use fractional counts in WoS categories and we consider that a paper reclassified in one of Topics[8] (Table 4).

Table 4. Modularity for three taxonomies (Corpus Y, period 2010-2022, full counts for links between WoS categories)

| | # Categories | # Links | H | H sdt error | Confidence Interval 99% | |
|---|---|---|---|---|---|---|
| OST | 242 | 398,776,924 | 0.30 | 0.04 | 0.20 | 0.40 |
| WoS | 244 | 1,217,037,695 | 0.08 | 0.01 | 0.06 | 0.10 |
| Leiden meso topics | 326 | 270,406,570 | 0.33 | 0.02 | 0.28 | 0.38 |

The OST modularity index was expected to be higher for OST classification compared to the WoS modularity due to its enhanced consistency between documents and their references. Furthermore, the OST modularity index is very similar to the modularity index of Leiden meso-topics. This result is very satisfactory since the Leiden classification maximizes this criterion for the definition of micro-topics.

*Accuracy*

Another simple index is used by Klavans and Boyack (2017) to compare the accuracies of various scientific taxonomies. Klavans and Boyack compare a classification of the scientific literature with reference sets of articles with very extensive bibliographies. They argue that such articles - usually review papers - frequently provide a summary of a research topic. The sets of their references can be "considered as expert-based partitions of the literature".

The concentration of these sets in a given taxonomy is thus proposed as an indicator of the taxonomy accuracy. They use the Herfindahl index of the references of a paper $s$

$$HE(s) = \sum_i p_i(s)^2$$ where $p_i(s)$ is the proportion of references of paper $s$ in class $i$.

The *accuracy* of a classification is then measured by the mean of the concentrations of a set of *gold standard* papers

$$HE = \frac{1}{n} \sum_s HE(s).$$

Following Klavans & Boyack (2017), gold standard papers are selected as papers with at least 100 references in the database.

Computing the mean Herfindahl indexes of the references of 129,019 gold standard papers - ignoring the references in removed WoS categories for the WoS index - shows that the accuracy of the OST reclassification is significantly higher than the WoS accuracy and that it is very similar to that of the Leiden classification into meso level topics (Table 5).

Table 5. Concentration of Golden Standard papers for three classifications (Corpus Z, period 2010-2022)

| Classification | # Categories | # Gold St | HE | HE(s) sdt error | HE Sdt error | Confidence Interval 99% | |
|---|---|---|---|---|---|---|---|
| OST | 242 | 129,019 | 0.473 | 0.225 | 0.001 | 0.472 | 0.475 |
| WoS | 244 | 129,019 | 0.204 | 0.133 | 0.000 | 0.203 | 0.205 |
| Leiden meso topics | 326 | 129,019 | 0.458 | 0.247 | 0.001 | 0.457 | 0.460 |

The OST classification exhibits a highly significant improvement from the WoS with this second index. Again, the OST index is similar to the Leiden meso-level index [9].

## 2 Exploring migrations from WoS to OST categories

As the overall rate of the OST classification revision is about 50%, It is instructive to look at how the categories are being

---

8 Data for *Leiden citation topics* as reported in the WoS database
9 Though a larger number of classes tends to decrease the concentration, the difference in class numbers between Leiden and OST classification does not have a significant impact on the difference between the two indexes. A linear transformation of the Leiden index in order to have the same variation range as the OST index would not change this modified. When a paper from a WoS category $C_1$ is reclassified in a different OST category $C_2$, we say that the paper *migrates* from $C_1$ to $C_2$.

The statistics we provide are in corpus Y for the period of 2010-2022.

## 2.1 Migrations between categories

Attention could be focused on two types of categories: attracting categories that gather many papers from other J-categories and dispersed categories when many papers are reclassified in other P-categories. We use the following ratios similar to those of Shu et al. (2020).

*Ratio J* is the percentage of papers of a J-category that are classified in a different P-Category where we use fractional counts for documents in multiple WoS categories.

*Ratio P* is the percentage of papers of the P-category that come from a different J-category where we only count papers that have not the OST category as one of its WoS category.

When Ratio J is high, the J-category is dispersed among other P-categories. When Ratio P is high, the P-category attracts documents from other J-categories. A category is deeply modified if both ratios are high: 47% of the categories are in the upper right quadrant on Figure 1.

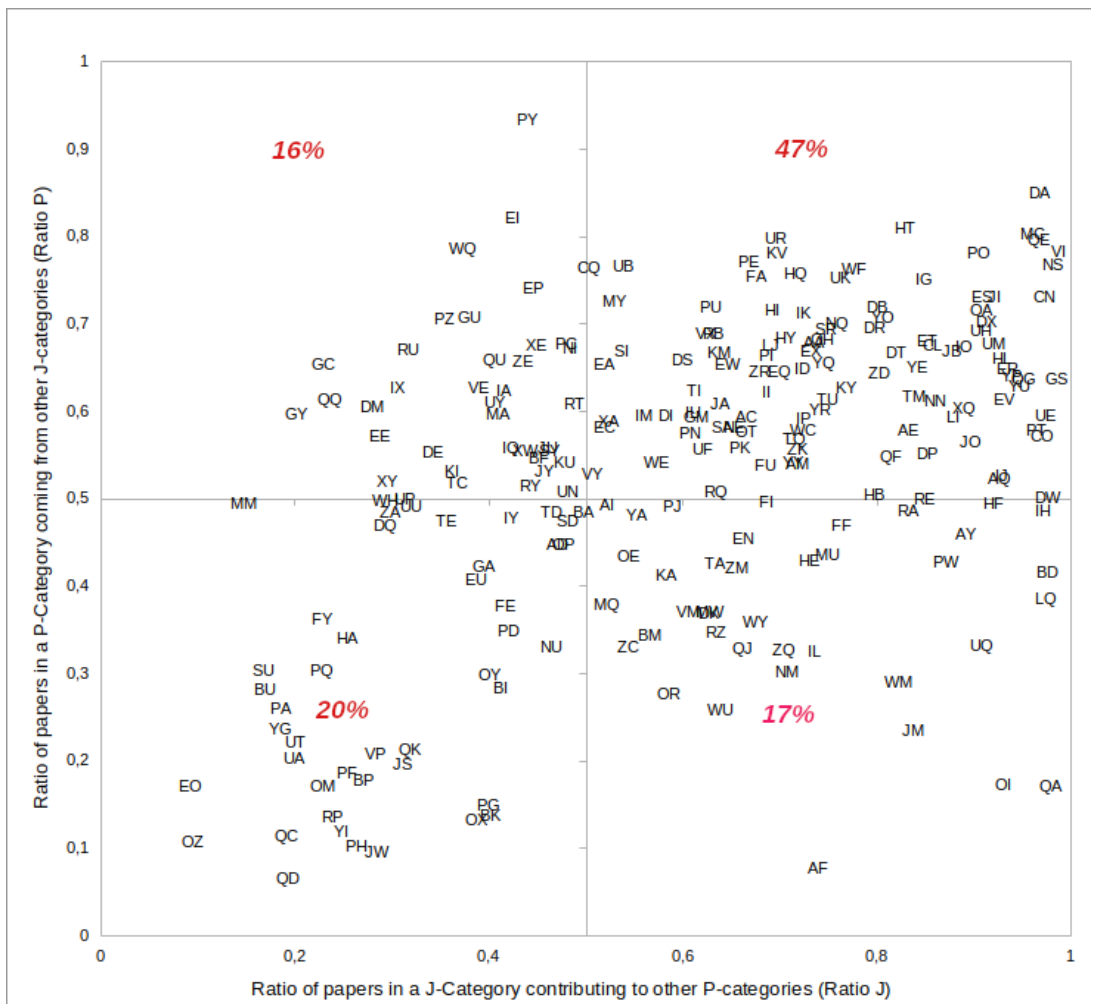

**Figure 1.** Comparison between journal classification and paper classification, by categories (2010-2022).

BIOPHYSICS (DA), MATHEMATICAL & COMPUTATIONAL BIOLOGY (MC), MATERIALS SCIENCE, BIOMATERIALS (QE), PSYCHOLOGY (VI), NANOSCIENCE & NANOTECHNOLOGY (NS), MATHEMATICS, INTERDISCIPLINARY APPLICATIONS (PO), BEHAVIORAL SCIENCES (CN) are among the deeply modified categories.

---

index before the third decimal.

In medical research, there is a rather loose relationship between the WoS medical specialities of journals and those of the papers published in those journals. Figure 1a shows that most OST categories have more than 50% of papers coming from another WoS category. The category MEDICINE, GENERAL & INTERNAL is a extreme case as 93.3% of papers in the OST category come from another WoS category, almost all of them in Medical Research (Table A1 in Appendix A).

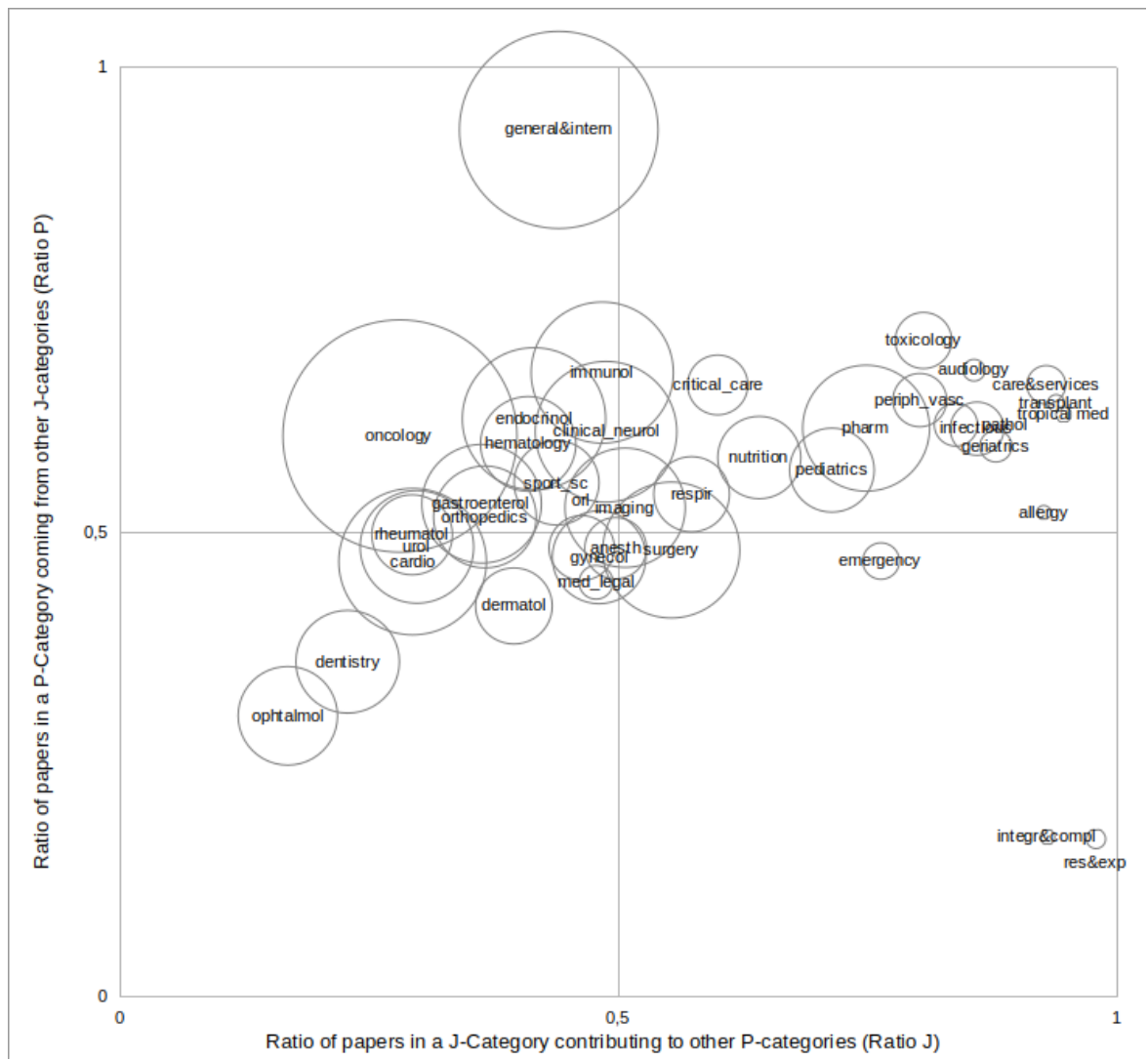

**Figure 1a.** Comparison between journal classification and paper classification for Medical categories (2010-2022). Bubble size is proportional to the number of documents in the OST category.

Many categories in the Humanities discipline are conserved, partly because the documents with less than two references in the WoS base remain in their J-category (Fig. 1b). The perimeters of Psychology categories have been extensively modified, which is related to their connections with Life sciences.

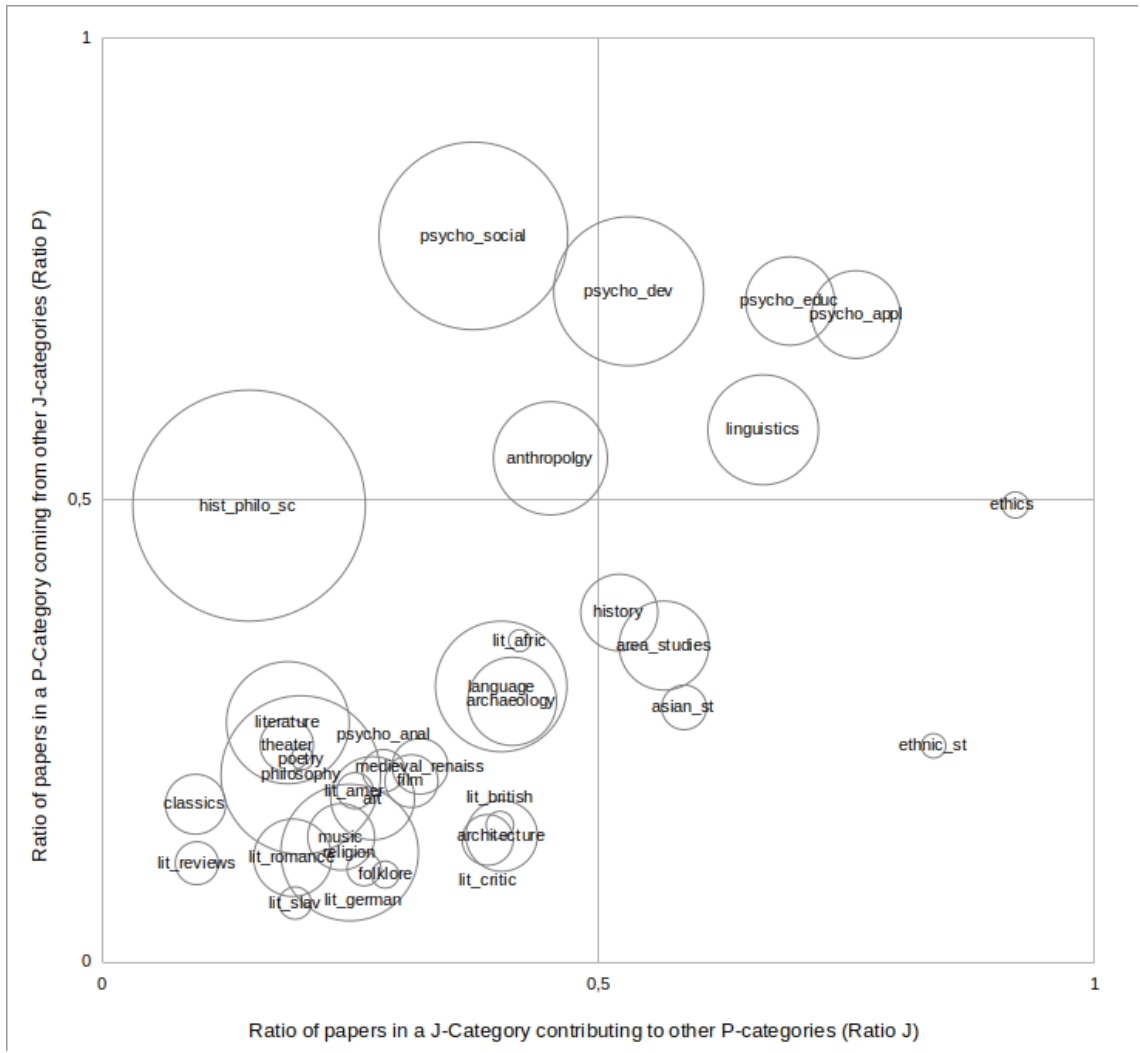

**Figure 1b.** Comparison between journal classification and paper classification for categories of the Humanities discipline (2010-2022). Bubble size is proportional to the number of documents in the OST category.

An alluvial graph showing migrations between all categories would be useful to describe these category reorganizations more accurately. As such a large graph is difficult to analyse, we limit the description to migrations between disciplines.

### 2.2 *Distribution of world publications into disciplines*

Migrations between categories affect disciplines when they happen between categories that belong to different disciplines. The overall distribution across disciplines shows that discipline sizes are roughly preserved (Figure 2). The main changes concern Medical Research and Biology which have increased by +16.2% and +28.1% respectively. The size of other disciplines has been reduced (Applied Biology, -11.9%, Engineering -9.4% and Computer Science -16.4% ).

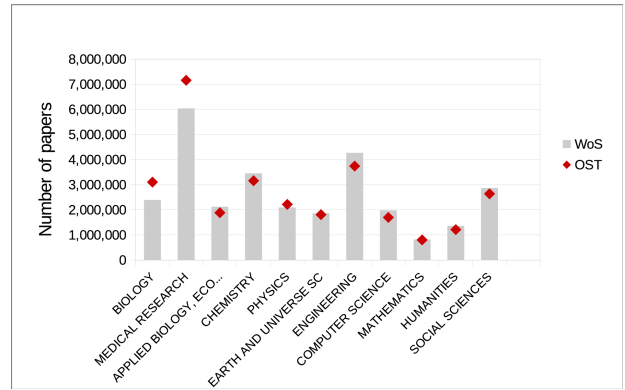

**Figure 2.** World distribution of papers across disciplines according to WoS and OST classifications for the publication period 2010-2022.

The alluvial graph of migrations (Figure 3) shows many exchanges between the three disciplines in the Life science domain and also between Chemistry and Physics. The Engineering discipline is partly redistributed in Chemistry, Physics and Computer Science disciplines. This can be explained as research papers in Engineering journals using knowledge or skills from chemistry, physics or computer science.

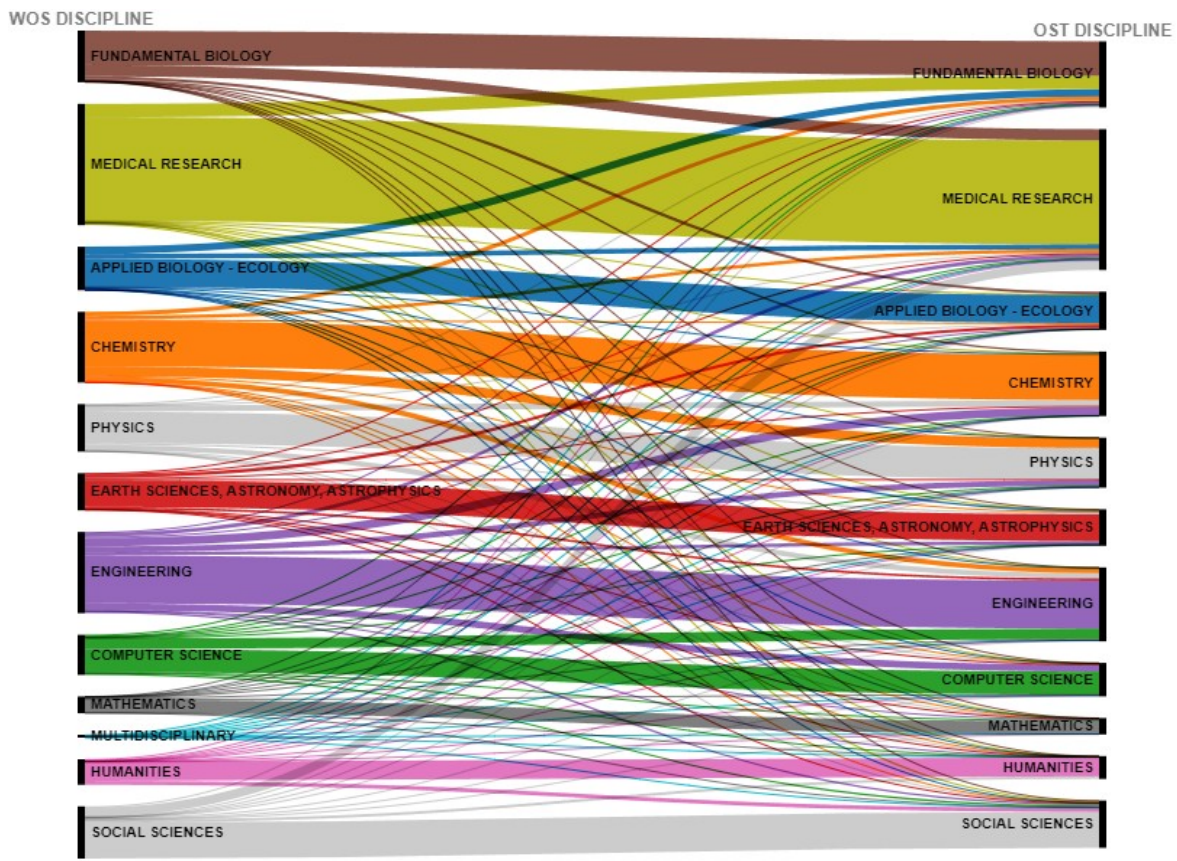

**Figure 3.** Alluvial graph of migrations from disciplines based on J-categories - denoted *WoS Disciplines* - to disciplines based on P-categories - denoted *OST Disciplines* - for the whole world (Corpus Y, period 2010-2022, *DataSet2, https://zenodo.org/records/15606281).*

Many papers that are originally published in Social Sciences journals are then reclassified as Medical Research. The division of certain categories between Social Sciences and Medical Research disciplines by OST (as indicated by the bold categories in Table 6) contributes to this shift.

**Table 6.** Migration counts from WoS Social Science categories to OST Medical Research discipline

| WoS Category | Code | % in (Soc Sc→ Med) migrations |
|---|---|---|
| **PUBLIC, ENVIRONMENTAL & OCCUPATIONAL HEALTH** | **NE** | 31.50 |
| **PSYCHIATRY** | **VE** | 12.30 |
| **NURSING** | **RZ** | 10.55 |
| HEALTH POLICY & SERVICES | LQ | 8.33 |
| **REHABILITATION** | **WC** | 6.28 |
| PSYCHOLOGY, CLINICAL | EQ | 4.33 |
| **GERONTOLOGY** | **LJ** | 3.34 |
| EDUCATION & EDUCATIONAL RESEARCH | HA | 3.31 |
| **SUBSTANCE ABUSE** | **GM** | 2.37 |
| SOCIAL SCIENCES, BIOMEDICAL | WV | 2.34 |
| SOCIAL SCIENCES, INTERDISCIPLINARY | WU | 2.32 |
| HOSPITALITY, LEISURE, SPORT & TOURISM | MW | 1.44 |
| SOCIAL WORK | WY | 1.15 |
| ECONOMICS | GY | 1.07 |
| | | 90.64 |

## 3 Impact of the reclassification on country specialization indexes

The specialization index of a country is the ratio of the share of papers in that country in the discipline to the same share in the whole world. When classification is revised, specialization indexes are modified because documents migrate from WoS to OST disciplines at different rates depending on the country. We report the ratio between the two specialization indexes (Fig. 4).

Countries with highest or lowest ratios mainly include Eastern countries. For example, some ratios have increased by more than 20% in China, Korea, Russia, Taiwan, and Pakistan. Some other ratios have decreased by more than 20% in Japan, Korea, Taiwan and Pakistan (Fig. 4). Conversely, changes for European countries and Australia are less than 10%. This means that these countries more often publish papers in journals of the "right" discipline, that is in a journal consistent with their references. This observation indicates that journal scopes may be more aligned with western publishing choices.

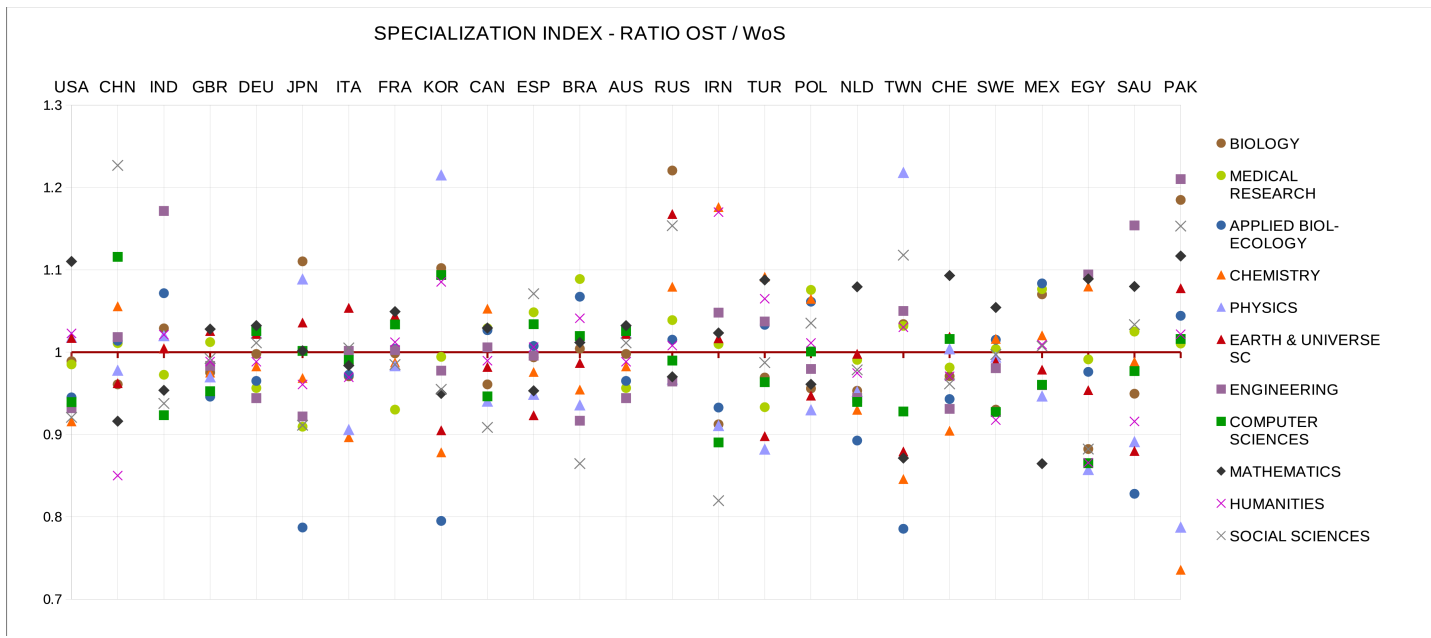

**Figure 4.** Ratio of specialization indexes for the 25 countries with more than 150,000 publications (2010-2022). Country order is of decreasing total number of publications. (*DataSet3*, https://zenodo.org/records/15606281).

To explore some of these variations, we use an approximation of the ratio $\rho$ of the specialization indexes as a function of the number of documents of three sets :

- the set *WoS(Disc)* of documents in the WoS discipline *Disc,*

- the set *NewOST(Disc)* of documents in the OST Discipline *Disc* that come from other WoS disciplines,

- the set *OldWoS(Disc)* of documents that were in the WoS discipline *Disc* and are reclassified in other OST disciplines.

This approximation (see Appendix B) is

$$\rho = 1 + [A(Disc, Country) - A(Disc, World)] - [B(Disc, Country) - B(Disc, World)] + \epsilon \qquad (1)$$

where

$$A(Disc, Country) = \frac{\#NewOST(Disc, Country)}{\#WoS(Disc, Country)} \quad \text{and}$$
$$B(Disc, Country) = \frac{\#OldWoS(Disc, Country)}{\#WoS(Disc, Country)}.$$

This means that the ratio depends on the difference of two rates between the country and the whole world: the attraction rate of (new) documents in the OST discipline and the loss rate of (old) documents from the WoS discipline.

Formula (1) explains the specialization ratio values represented in Figure 4 and provides additional information. It indicates whether a high or low value of the ratio is due to documents entering the OST discipline (Component A) or/and to documents exiting the WoS discipline (component B).

We show how to use this information in two examples.

### 3.1  Example 1: Country specialization in Applied Biology-Ecology

The low specialization ratios for Japan, Korea and Taiwan displayed in Figure 4 are mainly due to the high rate of papers leaving the WoS Applied Biology discipline, recorded in the B component (Table 7).

**Table 7.** Specialization ratio for Applied Biology (03) in four selected countries

| Country | # NewOST(03) | # OldWoS(03) | # WoS(03) | # NewOST/ #WoS = A | # OldWoS/ #WoS = B | A(Country) - A(World) | B(Country) - B(World) | A-B+1 = Ratio + o |
|---|---|---|---|---|---|---|---|---|
| JAPAN | 14,245.7 | 36,361.54 | 73,483.88 | 0.19 | 0.49 | **-0.08** | **0.11** | 0.81 |
| KOREA | 10,476.0 | 24,248.79 | 46,867.38 | 0.22 | 0.52 | -0.05 | **0.13** | 0.82 |
| TAIWAN | 4,101.7 | 8,475.84 | 14,467.26 | 0.28 | 0.59 | 0.01 | **0.20** | 0.81 |
| BRAZIL | 27,728.6 | 34,755.65 | 134,909.64 | 0.21 | 0.26 | **-0.07** | -0.13 | 1.06 |
| WORLD | 574,714.9 | 812,085.38 | 2,121,009.45 | 0.27 | 0.38 | 0.00 | 0.00 | 1.00 |

For these three countries, papers published in Applied Biology journals are more often (than the whole world) reclassified as Biology or Medical research works (Figure 5). In contrast, Brazil has fewer papers published in WoS Applied Biology journals that are reclassified as Biology or Medical Research works by OST.

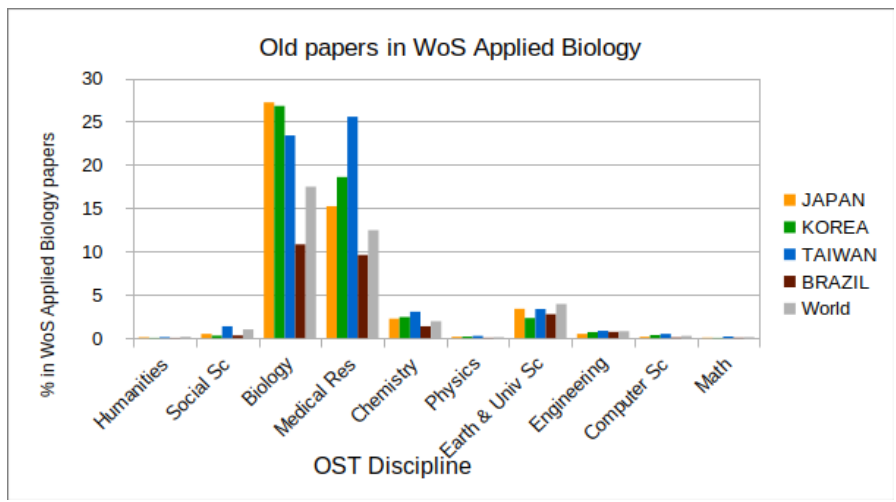

**Figure 5.** Percentages of papers that leave the WoS Applied Biology discipline for another OST discipline.

Considering new papers in the Applied Biology OST discipline (Figure 6), Brazil has lower migrations rates from Biology, Chemistry, Earth & Universe Sciences. This means that Brazilian papers in Applied Biology are less frequently published in journals of other disciplines. Conversely, this is not true for Korea, Taiwan and Japan.

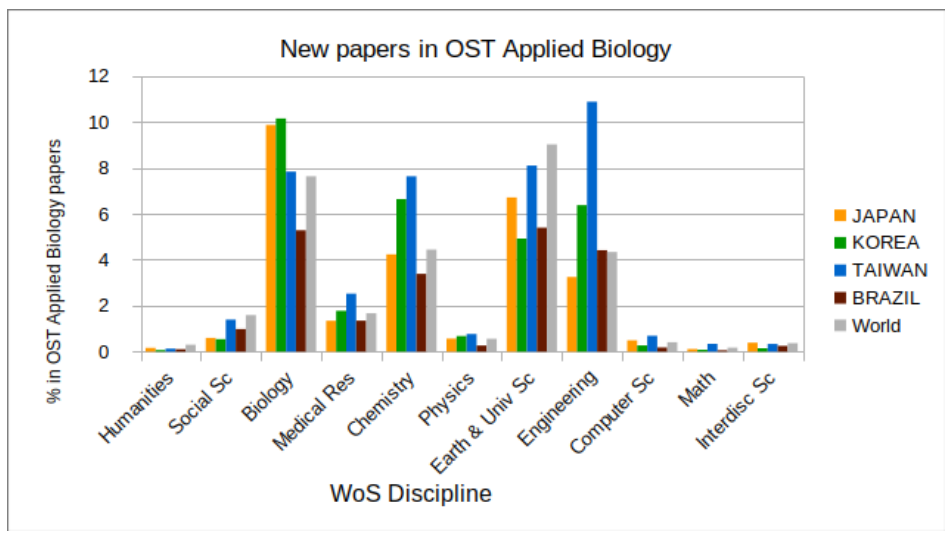

**Figure 6.** Percentages of papers reclassified in OST Applied Biology coming from other WoS disciplines

The two components of the specialization ratio show that Brazilian scientists choose Applied Biology journals to publish Applied Biology works with a high recall (low A component) and a high precision (low B component).

This may be related to Brazil's high specialization in Applied Biology (Fig. B1 & B3 in Appendix B).This high specialization in Applied Biology stems from the position and role of EMBRAPA[10], the Brazilian agricultural research institution supervised by the Department of Agriculture, Livestock and Food Supply (MAPA). The mission of the institution is to support agricultural production and agribusiness development, in a tropical environment. Some other research themes are explored including small farms, biodiversity conservation, and societal issues with an investment depending on the political context (Ollivier et al. 2019). As there are many Brazilian Applied Biology journals in the WoS database (Table B1 in the Appendix B), Brazilian scientists have full access to journals in the field.

---

10  Empresa Brasileira de Pesquisa Agropecuária

## 3.2 Example 2: Country specialization in Social Sciences

Canada, Brazil and Iran have low specialization ratios in Social Sciences. The *OldWoS* component (labelled B) is predominant for Canada and Iran (Table 8). This mainly corresponds to papers in Medical research journals reclassified in Social Sciences (Figure 7).

**Table 8.** Specialization ratio for Social Sciences in four selected countries

| Country | # NewOST(SS) | # OldWoS(SS) | # WoS(SS) | # NewOST/ #WoS = A | # OldWoS/ #WoS = B | A(Country) - A(World) | B(Country)- B(World) | A-B+1 = Ratio + o |
|---|---|---|---|---|---|---|---|---|
| IRAN | 7,068.0 | 13,124.9 | 24,641.9 | 0.29 | 0.53 | 0.05 | **0.22** | **0.83** |
| CANADA | 22,498.2 | 40,039.7 | 106,952.0 | 0.21 | 0.37 | -0.02 | **0.06** | **0.92** |
| BRAZIL | 10,393.6 | 28,059.5 | 86,408.1 | 0.12 | 0.32 | **-0.11** | 0.01 | **0.88** |
| CHINA | 77,376.1 | 51,527.8 | 200,841.5 | 0.39 | 0.26 | **0.15** | -0.06 | **1.21** |
| WORLD | 664,703.2 | 894,063.6 | 2,866,640.4 | 0.23 | 0.31 | 0.00 | 0.00 | 1.00 |

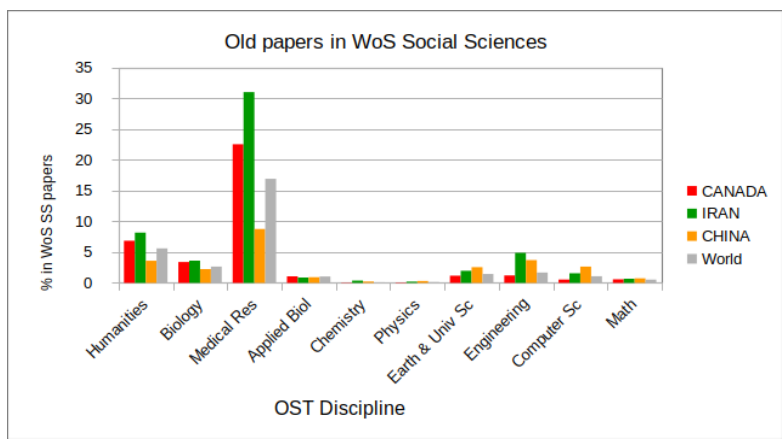

**Figure 7.** Percentages of papers leaving WoS Social Sciences for another OST discipline.

On the contrary, Brazil low ratio of specialization indexes is explained by a low value of the *New OST* Social Sciences component. There are less papers reclassified in OST Social Sciences coming from WoS Humanities and from other disciplines as Earth & Universe Sciences, Engineering or Computer sciences (Figure 8). Brazilian scientists less frequently publish Social Sciences works in journals of other disciplines.

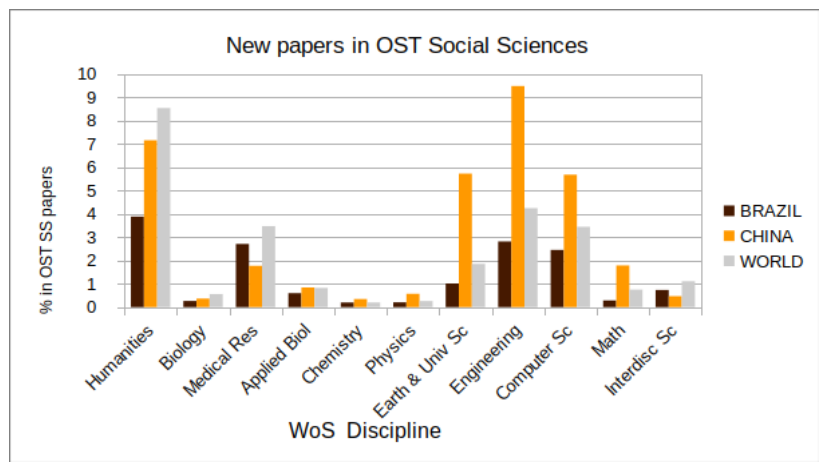

**Figure 8.** Percentages of OST Social Sciences papers published in journals of other WoS disciplines for Brazil and China.

The case of China high increase of the specialization ratio is related with many Social Sciences papers published in journals of other disciplines as Earth & Universe sciences (06), Engineering (07), Computer science (75) and Mathematics (Fig. 8).

Breaking down the N*ewOST(Social Sciences)* set into OST categories shows that the new papers arriving in Social Sciences - more frequently than the world - are in ECONOMICS, MANAGEMENT, BUSINESS and SOCIAL SCIENCES, INTERDISCIPLINARY (Fig. 9).

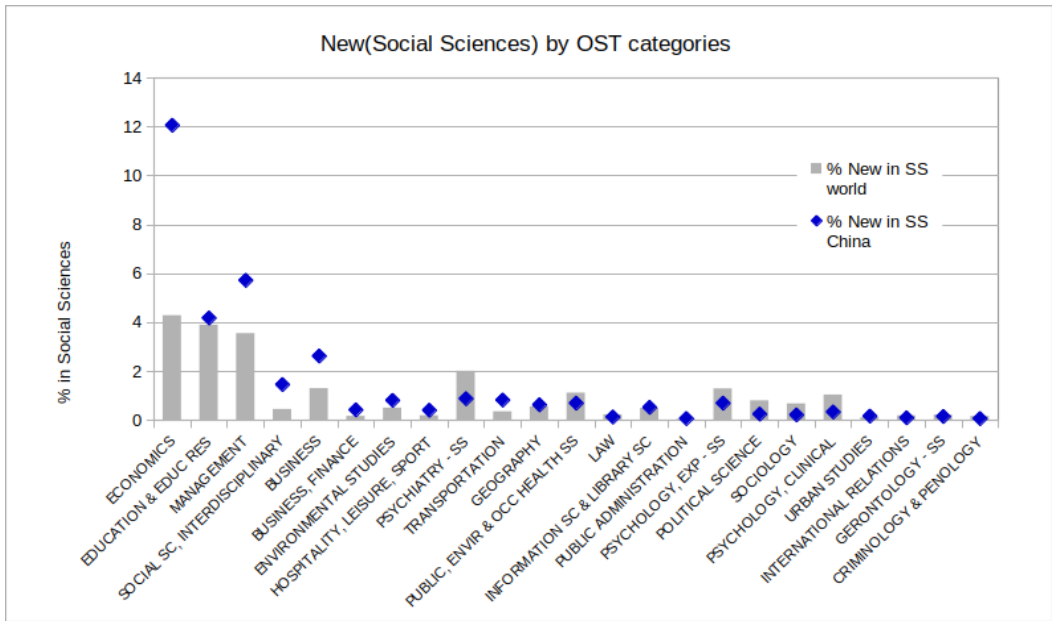

**Figure 9.** Distribution of China *new papers* in OST Social Sciences by categories (2010-2022).

China Social Sciences research is focused on a few categories (Fig. 10). In the OST classification the three four categories represent 72.4% of the Social Sciences in China when they are only 50.4% for the whole world.

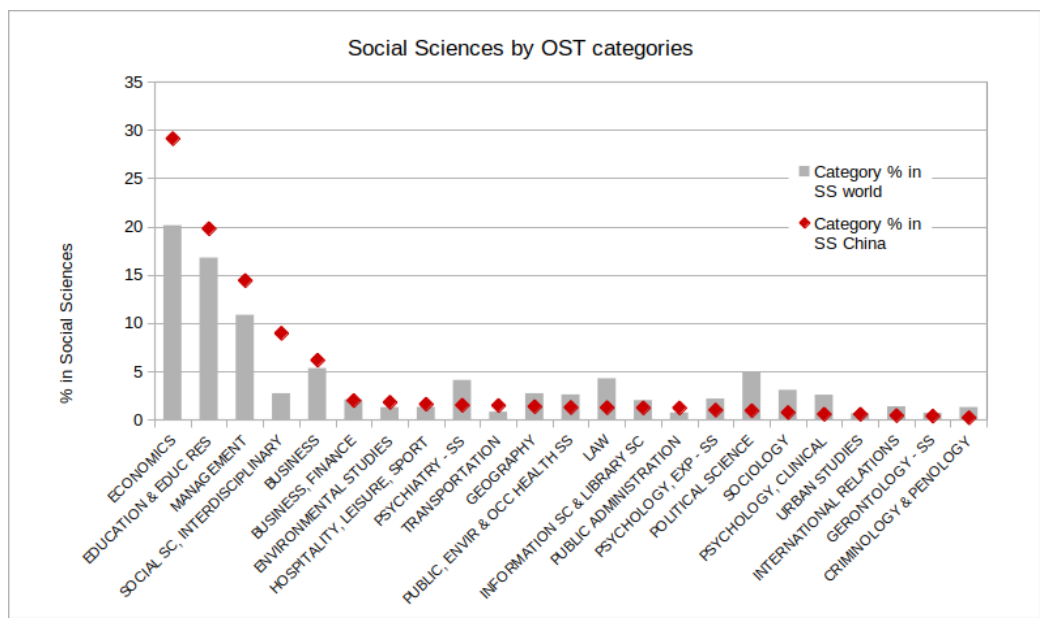

**Figure 10.** Distribution of China Social Sciences papers in OST categories (2010-2022).

In China, researchers have to navigate between academic requirements and political constraints. China's strategic priorities in economics research are focussed on topics such as economic growth, innovation, the impact of economic policies on national development and the development of 'the economic theory of socialism with Chinese characteristics' as explained in Zhang (2023). Sensitive topics such as political sociology remain taboo. The debate about the Chinese characteristics of methods and theories does not facilitate publication in international sociology journals (Merle, 2007; Breffeil & Dreyfuss, 2018).

The high rate of research in economics and sociology that is not published in international journals in these disciplines may be a result of this constrained research context. Is it because editorial committees are hesitant to accept these works or because researchers choose to publish their sociological work in other journals? The question of the publication strategies for researchers adapting to an authoritarian environment is raised by this example. However it is outside the scope of this study.

### 3.3 Interpreting changes in specialization

In this brief investigation, we discovered two countries that have distinct publishing strategies: Brazil scientists, which consistently publish their works in Applied Biology and Social Sciences in journals in those disciplines. In contrast, China scientists publish many of its Social Sciences works in journals that are not Social Sciences journals, at least for publications in journals selected by the WoS.

In general, countries that are not European nor North American or Australian have one or two disciplines where some specialization indices differ by more than 10% between the two classifications. This reflects a greater gap between the disciplines of their articles and those of the journals where they are published. This effect could be a result of a difference in scientific themes and methods between the country researchers and the editorial committees of the journals in the database.

These examples suggest interpretations that should be validated with more accurate data and analysis. The objective here was to show that variations in disciplinary specialization indicators reflect different choices of journals. These choices may differ depending on the academic positions and on research strategies of research performers.

## 4   Impact of the reclassification on MNCS indicators

Normalizing citation counts by field and year of publication as in the Normalized Citation Score (NCS) and its scale independent version MNCS (Waltman et al. 2010) is fairer than using citation counts. The database and its classification into fields affects the value of the related field-normalized indicators (Scheidsteger at al. 2023; Thelwall & Jiang, 2024).

We briefly examine the case of MNCS based on WoS or OST categories, averaged by discipline for the 25 countries, calculated using year 2019 data. The overall MNCS (all disciplines) has only been slightly modified, whereas more significant changes have been observed in disciplines as Physics, Engineering, Social Sciences, Humanities (Fig. C1 in the Appendix C) and Mathematics (Fig. 11).

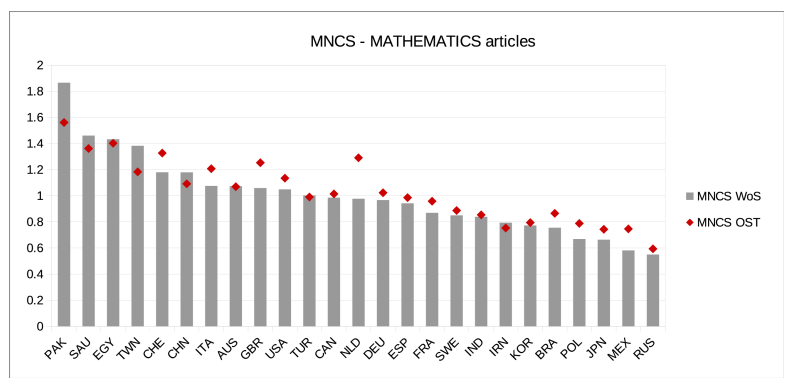

**Figure 11.**  MNCS Mathematics for WoS and OST normalizations (articles only, publication year 2019).

### 4.1   Breaking down the MNCS difference into intra- and inter-discipline components

We briefly explore the case of Mathematics to examine some issues related to this indicator.

The sum of normalized citation scores of the documents in a discipline changes

- when documents that are in the discipline for both classifications have different normalizing factors so that a same number of citations results in different normalized scores in the two classifications. These score differences contribute to the *intra-discipline* component of the difference of MNCS indicators;

- when documents are coming from one of the two sets of documents *NewOST(Math)* and *OldWoS(Math)* that are in the discipline for one classification and not for the other. Their contributions depend on the number of documents in these sets and on the difference between the scores of the documents in the two sets. They define the *inter-discipline* component of the MNCS difference (Formulas in Appendix C). Figure 12 displays the Mathematics MNCS difference broken down by intra- and inter-discipline components for the 25 countries.

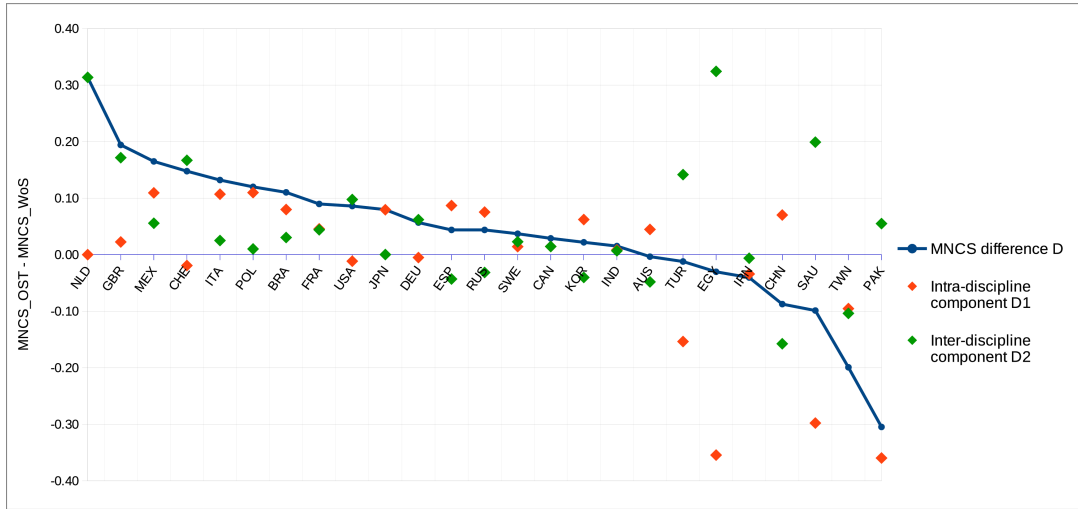

**Figure 12.** Difference of MNCS between WoS and OST for the Mathematics discipline and its decomposition into intra- and inter-discipline components (year 2019, articles only).

The intra-discipline MNCS variation is due to migrations between Mathematics categories with different mean number of citations. The main differences are between MATHEMATICS (PQ) and MATHEMATICS, APPLIED (PN) categories (Table 9).

**Table 9**. Mean number of citations (the normalizing denominator) in Mathematics categories for WoS and OST classifications

| Code | Category | WOS | OST |
|---|---|---|---|
| PN | MATHEMATICS, APPLIED | 6.81 | 8.49 |
| PQ | MATHEMATICS | 3.83 | 3.54 |
| XY | STATISTICS & PROBABILITY | 6.85 | 7.09 |
| PO | MATHEMATICS, INTERDISCIPLINARY APPLICATIONS | 8.90 | |

Therefore a high rate of migration from WoS-PN to OST-PQ contributes to a positive intra-discipline MNCS variation and a high rate from WoS-PQ to OST-PN contributes to a negative intra-discipline MNCS. In our sample, countries with the highest migration rates from WoS-PN to OST-PQ are Italy, Poland, Spain, Brazil and China which are among those with the highest intra_discipline component (Fig. 13). Countries with high rates from WoS-PQ to OST-PN are among the countries with the lowest intra-discipline component.

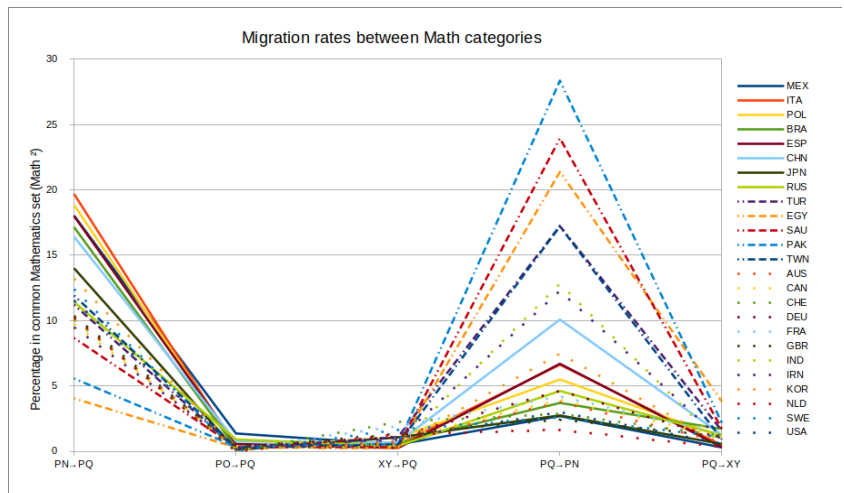

**Figure 13.** Main migrations between Mathematics WoS and OST discipline (articles only, year 2019).

In this example, these two migration rates explain well the value of the intra-discipline component of MNCS difference. In the general case, other migrations between Mathematics categories of highly cited papers could also significantly contribute to this difference.

The inter-discipline component is a balance between the contributions DA and DB of the two sets *NewOST(Math)* and *OldWoS(Math)* (Fig. C1 in the Appendix C). A positive inter-discipline component is associated with high scores of the *NewOST(Math)* set of papers. These documents come from different non Mathematics disciplines. Their scores depend on the OST category where they arrive but also on their citation numbers but not on the mean citation number of the category they come from. It is then necessary to identify the papers with the highest citation numbers in order to explain a large *NewOST(Math)* contribution. Similarly, individual high scores of documents in *OldWoS(Math)* have to be identified when a high DB component leads to a low value of the inter-discipline component. Such an analysis is therefore specific for each research performer.

*4.2 Discussion on MNCS sensitivity*

The MNCS indicator has been carefully designed to integrate into a single indicator publications from heterogeneous disciplines and ages. In the case of Mathematics, for example, it is appropriate to distinguish between pure mathematical papers that advance the discipline and those that utilize mathematics as an analytical tool for research in other disciplines. These two research activities have distinct communities of potential readers with varying sizes and citation practices. However, understanding why the MNCS varies when changing the classification is laborious and returning to the citation numbers of individual papers may be necessary.

The brief analysis above confirms that this indicator is sensitive to the choice of classification. It is therefore safer to use this indicator with the most relevant available classification. In OST standard reports, we assume that, since 2023, field normalized indicators have been based on these modified categories.

# 5 Conclusion

The OST reclassification of scientific publications in modified WoS categories meets our first requirements for a scientific nomenclature: a classification at paper level, with no overlapping of categories and no multidisciplinary categories. It also restores a better consistency between each paper and its references and improves the overall accuracy as compared to the references of gold standard papers. The new categories have the same names as the WoS categories and this is convenient for the regular recipients of OST reports. However, the familiarity with WoS category names may be misleading because some category perimeters have significantly changed. OST is currently updating the category descriptions by supplying sets of keywords to replace the previous descriptions of the WoS categories. Programming the reclassification is simple, requires minimal resources, and is easily updated annually. This is therefore a cheap method to approach the properties of direct, efficient classifications such as the Leiden classification (Waltman & van Eck, 2012, Traag et al. 2019). The method proposed by Milojević (2020) is therefore considered highly relevant and has been successfully adopted by OST.

Differences between journal and paper categories result in changes in country specialization indexes. Large discrepancies reveal unusual journal choices which are more common among non western countries. National scientists have either facilitated or constrained choices when selecting WoS journals. This is observed for Brazilian scientists who easily publish their works in Applied Biology or in Social Sciences journals. Chinese scientists use alternative publication strategies to publish all their works in economics and sociology.

MNCS is an elaborate indicator which is designed for large and diverse set of publications but is sensitive to the classification. Analysing particular changes of the indicator may be difficult as it does not depend only on migration rates between fields. This indicator is more reliable if fields are consistent with the referencing/citation usage of each research community.

This study aims to provide a preliminary outline of how to comprehend the effects of switching from a journal to a paper level classification on usual indicators. As OST produces regular reports for French universities and institutions, it is important to explain the major changes in these indicators. The examples proposed in this paper provide a method for analysing such institutions. This is necessary because indicators for highly specialized institutions can be significantly altered and, for those institutions, it may be useful to have specific interactions with OST to assist them in their self-assessment.

**Acknowledgments**

This work uses Web of Science data by Clarivate Analytics in the in-house version of Hcéres-OST.

We are grateful to Luis Miotti for his advice on experimenting with Milojević's method and for his support during the entire OST nomenclature revision process. Lesya Baudoin helped to make efficient decisions at the beginning of this project and the other colleagues of the OST team got involved during the testing phase.

We thank the two reviewers for their comments and recommendations which enabled us to deeply improve the initial version of this paper.

**Author contributions**

Agénor Lahatte: conceptualization, data curation, formal analysis, software, validation, writing review

Élisabeth de Turckheim: conceptualization, formal analysis, validation, visualization, writing.

**Competing interests**

No competing interests to declare.

**Funding**

This work was funded by the Hcéres-OST current budget.

**Data availability**

The data used in this paper is proprietary and cannot be posted in a repository. Intermediate data are given in Lahatte&Turckheim (2025) Reclassification_Data (https://zenodo.org/records/15606281).

# 6 References

Archambault, É., Beauchesne, 0. & H., Caruso, J. (2011) Towards a multilingual, Comprehensive and Open Scientific Journal Ontology. *Proceedings of the 13th International Conference of the International Society for Scientometrics and Informetrics ISSI* (pp. 66–77).
*http://www.science-metrix.com/pdf/Towards_a_Multilingual_Comprehensive_and_Open.pdf*
*https://zenodo.org/records/10030868*

Bassecoulard, E. & Zitt, M. (1999). Indicators in a research institute: A multi-level classification of scientific journals. *Scientometrics*, 44(3), 323–345. *https://doi.org/10.1007/BF02458483*

Börner, K., Klavans, R., Patek, M., Zoss, A. M., Biberstine, J. R., Light R. P., Larivière, V., Boyack, K. W. (2012). Design and Update of a Classification System: The UCSD Map of Science. *PLoS ONE* 7(7)
*https://journals.plos.org/plosone/article?id=10.1371/journal.pone.0039464*

Boyack, K. W., Newman, D., Duhon, R. J., Klavans, R., Patek, M., Biberstine, J. R., et al. (2011). Clustering more than two million biomedical publications: Comparing the accuracies of nine text-based similarity approaches. *PLoS ONE, 6*(3), e18029. *https://doi.org/10.1371/journal.pone.0018029*

Breffeil, E. & Dreyfuss, J. (2018). Faire de la sociologie avec un État autoritaire. Le cas de la Chine. *Sociologies pratiques,* 36(1), 121-130. https://doi.org/10.3917/sopr.036.0121.

CWTS Leiden Ranking (2024). *https://www.leidenranking.com/information/fields* ,
*https://www.leidenranking.com/information/indicators#impact-indicators*

Klavans, R., & Boyack, K. W. (2017). Which type of citation analysis generates the most accurate taxonomy of scientific and technical knowledge? *Journal of the Association for Information Science and Technology, 68*(4), 984–998. *https://doi.org/10.1002/asi.23734*

Leydesdorff, L., & Bornmann, L. (2016). The operationalization of "fields" as WoS subject categories (WCs) in evaluative bibliometrics: The cases of "library and information science" and "science & technology studies." *Journal of the Association for Information Science and Technology,* 67(3), 707–714. *https://doi.org/10.1371/journal.pone.0018029*

Leydesdorff, L., Bornmann, L. & Wagner, C. S. (2017) Generating Clustered *Journal Maps*: An Automated System for Hierarchical Classification *Scientometrics*. *https://doi.org/10.1007/s11192-016-2226-5*

Merle, A. (2007) De la reconstruction de la discipline à l'interrogation sur la transition : la sociologie chinoise à l'épreuve du temps. *Cahiers internationaux de sociologie* 122 *https://shs.cairn.info/revue-cahiers-internationaux-de-sociologie-2007-1-page-31?lang=fr*

Milojević, S. (2020). Practical method to reclassify Web of Science articles into unique subject categories and broad disciplines. *Quantitative Science Studies, https://doi.org/10.1162/qss_a_00014*

Ollivier, G., Bello, S., Deane de Abreu Sa, T., & Magda, D. (2019) The boundaries of agroecology. Research policies of two public agricultural institutes in France and Brazil. *Natures Sciences Sociétés* 27, 1, 20-38 *https://doi.org/10.1051/nss/2019017*

Ruiz-Castillo, J., & Waltman, L. (2015). Field-normalized citation impact indicators using algorithmically constructed classification systems of science. *Journal of Informetrics*, *9*(1), 102–117. *https://doi.org/10.1016/j.joi.2014.11.010*

Scheidsteger, T, Haunschield, R. & Bornmann, L. (2023) How similar are field-normalized scores from different free or commercial databases calculated for large German universities? *27th International Conference on Science, Technology and Innovation Indicators* (STI 2023) https://dapp.orvium.io/deposits/64f83f76269bc6dc8000269c/view

Shu, F., Julien, C.-A., Zhang, L., Qiu, J., Zhang, J. & Larivière, V. (2019). Comparing journal and paper level classifications of science. *Journal of Informetrics, 13(1), 202–225. https://doi.org/10.1016/j.joi.2018.12.005*

Shu, F., Ma, Y., Qiu, J. & Larivière, V. (2020). Classifications of science and their effects on bibliometric evaluations. *Scientometrics* **125**, 2727-2744 *https://doi.org/10.1007/s11192-020-03701-4*

Thelwall, M. & Fairclough, R. (2017). The accuracy of confidence intervals for Field normalized indicators. *J. of Informetrics* *https://arxiv.org/abs/1703.04031*

Thelwall, M. & Jiang, X. (2024) Is OpenAlex Suitable for Research Quality Evaluation and Which Citation Indicator is Best? *https://arxiv.org/abs/2502.18427*

Traag, V. A., Waltman, L. & vanEck, N. J. (2019). From Louvain to Leiden : guaranteeing well-connected communities. *Scientific Reports* *https://doi.org/10.1038/s41598-019-41695-z*

Waltman, L., van Eck, N.J., van Leeuwen, T. N., Visser, M.S. & van Raan, A. (2010). Towards a new crown indicator: Some theoretical considerations. *Journal of Informetrics* 5(1):37-47 *https://doi.org/10.1016/j.joi.2010.08.001*

Waltman, L. & van Eck, N. J. (2012). A new methodology for constructing a publication-level classification system of science. *J. Assoc Inf Sci Technol*. 66-12 (pp 2378-2392). *https://doi.org/10.1002/asi.22748*

Waltman, L., & van Eck, N. J. (2024) *An open approach for classifying research publications* *https://www.leidenmadtrics.nl/articles/an-open-approach-for-classifying-research-publications*

Wang, Q., & Waltman, L., (2016). Large-scale analysis of the accuracy of the journal classification systems of Web of Science and Scopus. *J. of Informetrics* *https://doi.org/10.1016/j.joi.2016.02.003*

Zhang, Z. (2023). Overview: Economic Research in China: Analysis Based on Papers Published in the Economic Research Journal. In: Handbook of Chinese Economics. Springer, Singapore. *https://link.springer.com/chapter/10.1007/978-981-99-0420-4_1*

# 7 APPENDIX A. Migrations to the OST category MEDICINE, GENERAL & INTERNAL (PY)

**Table A1.** 30 first WoS categories migrating to the OST- MEDICINE, GENERAL & INTERNAL category

| Code | WoS CATEGORY | % in OST-PY | Code | WoS CATEGORY | % in OST-PY |
|------|--------------|-------------|------|--------------|-------------|
| NE | PUBLIC, ENVIRONMENTAL & OCCUPATIONA | 10.87 | IA | ENDOCRINOLOGY & METABOLISM | 1.62 |
| HL | HEALTH CARE SCIENCES & SERVICES | 7.30 | PT | MEDICAL INFORMATICS | 1.40 |
| **PY** | **MEDICINE, GENERAL & INTERNAL** | **6.68** | LI | GERIATRICS & GERONTOLOGY | 1.36 |
| NN | INFECTIOUS DISEASES | 4.81 | WE | RESPIRATORY SYSTEM | 1.32 |
| TU | PHARMACOLOGY & PHARMACY | 3.95 | FF | EMERGENCY MEDICINE | 1.29 |
| DQ | CARDIAC & CARDIOVASCULAR SYSTEMS | 3.92 | SD | OBSTETRICS & GYNECOLOGY | 1.28 |
| LQ | HEALTH POLICY & SERVICES | 2.94 | DM | ONCOLOGY | 1.16 |
| RZ | NURSING | 2.51 | HB | EDUCATION, SCIENTIFIC DISCIPLINES | 1.09 |
| QA | MEDICINE, RESEARCH & EXPERIMENTAL | 2.44 | QU | MICROBIOLOGY | 1.08 |
| YA | SURGERY | 2.32 | VE | PSYCHIATRY | 1.05 |
| TQ | PEDIATRICS | 2.11 | DS | CRITICAL CARE MEDICINE | 0.94 |
| NI | IMMUNOLOGY | 2.10 | SA | NUTRITION & DIETETICS | 0.93 |
| ML | PRIMARY HEALTH CARE | 1.99 | ZA | UROLOGY & NEPHROLOGY | 0.90 |
| ZD | PERIPHERAL VASCULAR DISEASE | 1.92 | MA | HEMATOLOGY | 0.88 |
| RT | CLINICAL NEUROLOGY | 1.64 | BA | ANESTHESIOLOGY | 0.85 |
|  |  | 57.49 |  |  | 17.15 |

# 8 APPENDIX B. Specialization indexes

## 8.1 Country specialization by discipline

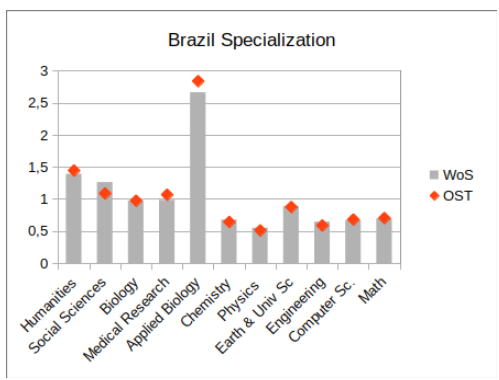

**Figure B1.** Brazil specialization in the 11 disciplines.

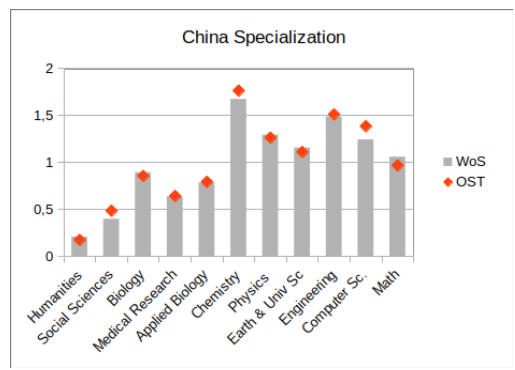

**Figure B2.** China specialization in the 11 disciplines.

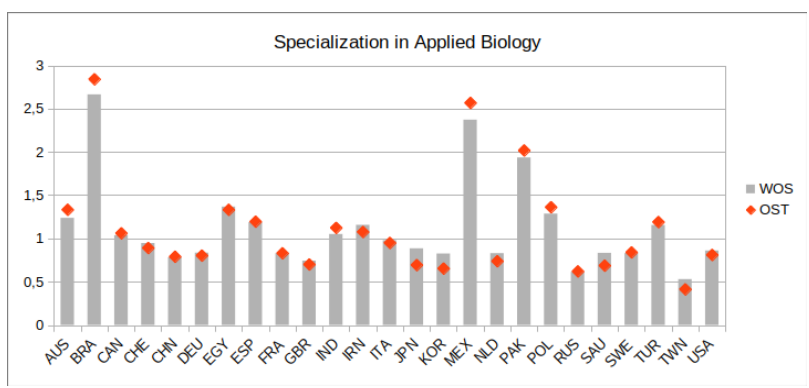

**Figure B3.** Country specialization in Applied Biology.

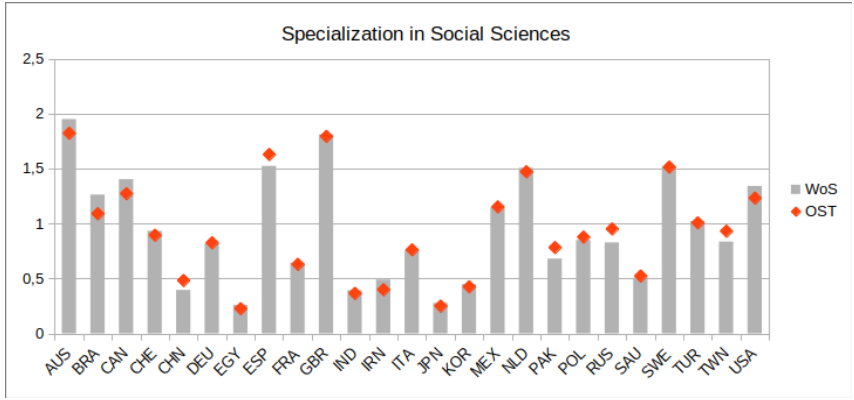

**Figure B4.** Country specialization in Social Sciences.

### 8.2 Approximation of the ratio of specialization indexes $\rho$

The ratio of the specialization indexes between OST and WoS $\rho$ is such that

$$\frac{\#OST(Disc,Country)}{\#WoS(Disc,Country)} = \rho \frac{\#OST(Disc,World)}{\#WoS(Disc,World)}$$

To relate this inequality with numbers of migrations, we define

$\#OST(Disc)$ as the total number of papers in the OST discipline *Disc*,

$\#NewOST(Disc)$ the number of papers in *OST_Disc* that come from another WoS discipline,

$\#OldWoS(Disc)$ the number of papers in *WoS_Disc* that leave to another OST discipline.

As $\#OST(Disc) = \#WoS(Disc) + \#NewOST(Disc) - \#OldWoS(Disc)$

$$\frac{\#OST(Disc)}{\#WoS(Disc)} = 1 + \frac{\#NewOST(Disc)}{\#WoS(Disc)} - \frac{\#OldWoS(Disc)}{\#WoS(Disc)}$$

Using, $\frac{1+x}{1+y} = 1 + x - y + o(x+y)$,

a first order approximation of the specialization ratio is

$$\rho = 1 + A(Country) - A(World) - (B(Country) - B(World)) + \epsilon$$

where

$$A(Country) = \frac{\#NewOST(Disc,Country)}{\#WoS(Disc,Country)} \quad \text{and} \quad B(Country) = \frac{\#OldWoS(Disc,Country)}{\#WoS(Disc,Country)}$$

## 8.3 Brazilian publications in Applied Biology

**Table B1.** The 100 first journals publishing Brazilian Applied Biology research (Period 2010- 2022)

| Journal | Count | % |
|---|---:|---:|
| ZOOTAXA | 3501 | 4.73 |
| SEMINA CIENCIAS AGRARIAS | 3145 | 4.25 |
| CIENCIA RURAL | 3097 | 4.18 |
| ARQUIVO BRASILEIRO DE MEDICINA VETERINARIA E ZOOTECNIA | 2133 | 2.88 |
| PESQUISA AGROPECUARIA BRASILEIRA | 1752 | 2.37 |
| BIOSCIENCE JOURNAL | 1719 | 2.32 |
| REVISTA BRASILEIRA DE ZOOTECNIA BRAZILIAN JOURNAL OF ANIMAL SCIENCE | 1667 | 2.25 |
| PESQUISA VETERINARIA BRASILEIRA | 1642 | 2.22 |
| PLOS ONE | 1566 | 2.12 |
| REVISTA BRASILEIRA DE ENGENHARIA AGRICOLA E AMBIENTAL | 1498 | 2.02 |
| REVISTA BRASILEIRA DE FRUTICULTURA | 1464 | 1.98 |
| PHYTOTAXA | 1421 | 1.92 |
| REVISTA BRASILEIRA DE CIENCIA DO SOLO | 1326 | 1.79 |
| REVISTA CAATINGA | 1250 | 1.69 |
| CIENCIA FLORESTAL | 1221 | 1.65 |
| ACTA SCIENTIAE VETERINARIAE | 1161 | 1.57 |
| FOOD SCIENCE AND TECHNOLOGY | 1142 | 1.54 |
| REVISTA CIENCIA AGRONOMICA | 1099 | 1.48 |
| ANAIS DA ACADEMIA BRASILEIRA DE CIENCIAS | 1045 | 1.41 |
| BIOTA NEOTROPICA | 1029 | 1.39 |
| HORTICULTURA BRASILEIRA | 1001 | 1.35 |
| FOOD RESEARCH INTERNATIONAL | 997 | 1.35 |
| REVISTA ARVORE | 984 | 1.33 |
| ACTA HORTICULTURAE | 923 | 1.25 |
| GENETICS AND MOLECULAR RESEARCH | 917 | 1.24 |
| LWT FOOD SCIENCE AND TECHNOLOGY | 915 | 1.24 |
| ACTA BOTANICA BRASILICA | 899 | 1.21 |
| FOOD CHEMISTRY | 879 | 1.19 |
| PLANTA DANINHA | 868 | 1.17 |
| CIENCIA E AGROTECNOLOGIA | 829 | 1.12 |
| BRAZILIAN JOURNAL OF BIOLOGY | 778 | 1.05 |
| SCIENTIA FORESTALIS | 759 | 1.03 |
| FLORESTA E AMBIENTE | 756 | 1.02 |
| ACTA SCIENTIARUM AGRONOMY | 752 | 1.02 |
| NEOTROPICAL ENTOMOLOGY | 716 | 0.97 |
| ENGENHARIA AGRICOLA | 655 | 0.88 |
| SCIENTIFIC REPORTS | 650 | 0.88 |
| SCIENTIA AGRICOLA | 641 | 0.87 |
| BRAGANTIA | 615 | 0.83 |
| REVISTA BRASILEIRA DE ENTOMOLOGIA | 609 | 0.82 |
| *Total in the 40 first journals* | 50,021 | 67.56 |
| *Total in the 100 first journals* | 74035.00 | 100.00 |

# 9 APPENDIX C. MNCS

## 9.1 Country MNCS for four disciplines

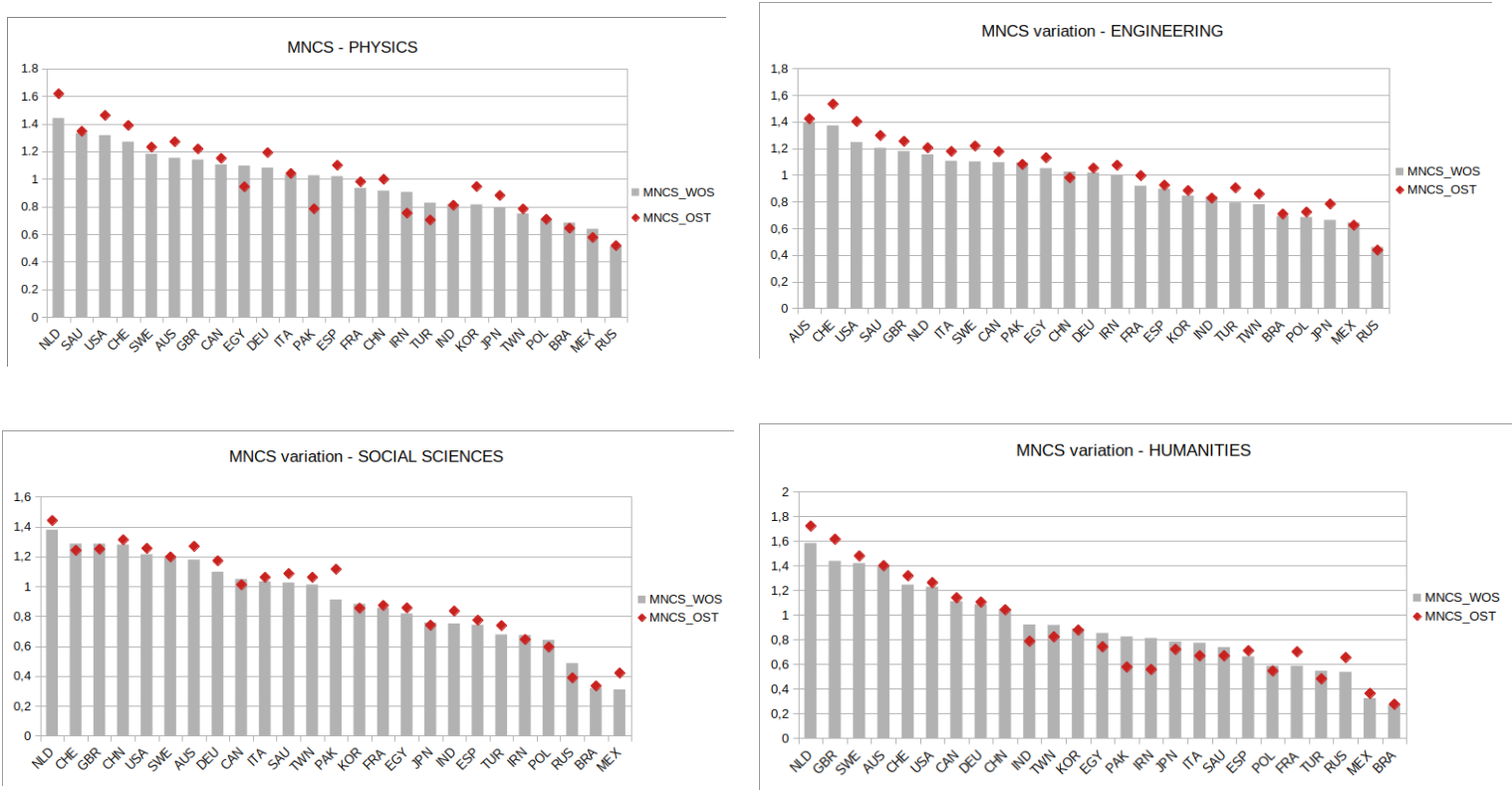

**Figures C1.** MNCS variation for 4 disciplines, 25 countries, year 2019.

## 9.2 Breaking down MNCS differences into intra- and inter-discipline components

*NewOST(Math)* is the set of documents in the OST discipline Mathematics from another WoS discipline, *OldWoS(Math)* are the documents in the WoS discipline Mathematics reclassified in another OST discipline. In *Math²* are the documents common to *WoS(Math)* and *OST(Math)*.

$$NCS_{OST}(OST(Math)) = NCS_{OST}(Math^2) + NCS_{OST}(NewOST(Math))$$

$$NCS_{WoS}(WOS(Math)) = NCS_{WoS}(Math^2) + NCS_{WoS}(OldWoS(Math))$$

$$D = MNCS_{OST}(OSTMath) - MNCS_{WoS}(WoSMath) = D1 + D2$$

$$D1 = \frac{NCS_{OST}(Math^2)}{\#OSTMath} - \frac{NCS_{WoS}(Math^2)}{\#WoSMath}$$  Intra-discipline component

$$D2 = DA - DB = \frac{NCS_{OST}(NewOST(Math))}{\#OSTMath} - \frac{NCS_{WoS}(OldWoS(Math))}{\#WoSMath}$$  Inter-discipline component

## 9.3 MNCS inter-discipline component

Decomposition of the Mathematics inter-discipline MNCS variation D2 into DA (*NewOST* contribution) and DB (*OldWoS* contribution)

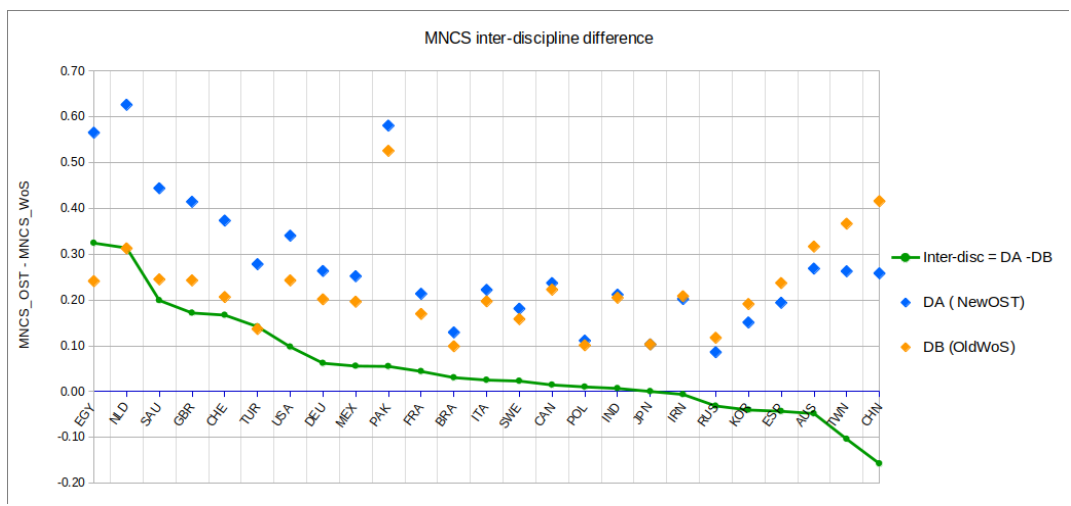

**Figure C2.** Decomposition of the inter-discipline variation for the MNCS in Mathematics for 25 countries.

To further explore the case of a country, it is necessary to pick up highly cited documents in the *NewOST* set and to check that they are more cited than those in their OST category. For example, Table A1 displays the WoS disciplines of the main categories arriving in *NewOST(Math)* for the two countries with the highest inter-discipline component. Exploring these documents is a good way to start the search for influential documents, which have a higher number of citations than the world in their OST category. For instance when there is a high DA component, one could search papers with mathematics references arriving from other discipline journals that are numerous and more cited than standard mathematics papers.

*Table C1.* Distribution of *NewOST(Math)* into WoS disciplines for The Netherlands and Egypt (fractional counts by address)

| Country | WoS Disc code | WoS Discipline | OST Cat code | OST category | # (WoS disc → OST Cat) | # *New Math* | % in *New Math* | % in *New Math* |
|---|---|---|---|---|---|---|---|---|
| NETHERLANDS | SS | SOCIAL SCIENCES | XY | STATISTICS & PROBABILITY | 33.14 | 156.86 | 21.12 | 61.49 |
| NETHERLANDS | 075 | COMPUTER SCIENCE | XY | STATISTICS & PROBABILITY | 14.70 | | 9.37 | |
| NETHERLANDS | SH | HUMANITIES | XY | STATISTICS & PROBABILITY | 14.19 | | 9.05 | |
| NETHERLANDS | 01 | BIOLOGY | XY | STATISTICS & PROBABILITY | 12.03 | | 7.67 | |
| NETHERLANDS | 02 | MEDICAL RESEARCH | XY | STATISTICS & PROBABILITY | 11.29 | | 7.20 | |
| NETHERLANDS | 07 | ENGINEERING | XY | STATISTICS & PROBABILITY | 11.10 | | 7.08 | |
| EGYPT | 05 | PHYSICS | PN | MATHEMATICS, APPLIED | 34.63 | 90.86 | 38.11 | 72.31 |
| EGYPT | 07 | ENGINEERING | PN | MATHEMATICS, APPLIED | 21.85 | | 24.04 | |
| EGYPT | 075 | COMPUTER SCIENCE | PN | MATHEMATICS, APPLIED | 9.22 | | 10.15 | |